\documentclass[twocolumn]{el-author}

\newcommand{\ua}{\uparrow}
\newcommand{\nc}{\newcommand}
\nc{\da}{\downarrow} \nc{\hc}{\hat{c}} \nc{\hS}{\hat{S}}
\nc{\bra}{\langle} \nc{\ket}{\rangle} \nc{\eq}{equation (\ref}
\nc{\h}{\hat} \nc{\hT}{\h{T}}\nc{\be}{\begin{eqnarray}}
\nc{\ee}{\end{eqnarray}}\nc{\rd}{\textrm{d}}\nc{\e}{eqnarray}\nc{\hR}{\hat{R}}\nc{\Tr}{\mathrm{Tr}}
\nc{\tS}{\tilde{S}}\nc{\tr}{\mathrm{tr}}\nc{\8}{\infty}\nc{\lgs}{\bra\ua,\phi|}\nc{\rgs}{|\ua,\phi\ket}
\nc{\hU}{\hat{U}}\nc{\lfs}{\bra\phi|}\nc{\rfs}{|\phi\ket}\nc{\hZ}{\hat{Z}}\nc{\hd}{\hat{d}}\nc{\mD}{\mathcal{D}}
\nc{\bd}{\bar{d}}\nc{\bc}{\bar{c}}\nc{\mc}{\mathcal}\nc{\ea}{eqnarray}\nc{\mG}{\mathcal{G}}\nc{\bce}{\begin{center}}
\nc{\ece}{\end{center}}
\newcommand{\overbar}[1]{\mkern 1.5mu\overline{\mkern-2.5mu#1\mkern-1.5mu}\mkern 1.5mu}
\date{12th December 2011}

\begin{document}

\title{Novel size effects on magneto-optics in the spherical quantum dots}

\author{Manvir S. Kushwaha}

\abstract{We embark on investigating the magneto-optical absorption in {\em spherical} quantum dots
{\em completely} confined by a harmonic potential and exposed to an applied magnetic field in the
symmetric gauge. This is done within the framework of Bohm-Pines' RPA that enables us to derive and
discuss the full Dyson equation that takes proper account of the Coulomb interactions. Intensifying
the confinement or magnetic field and reducing the dot-size yields a blue-shift in the absorption
peaks. However, the size effects are seen to be predominant in this role. The magnetic field tends to
maximize the localization of the particle, but leaves the peak position of the radial distribution
intact. The intra-Landau level transitions are forbidden.}

\maketitle

\section{Introduction}
The scientific quest behind the synthesis of semiconducting quantum dots is to create and control future
electronic and optical nanostructures, which are quantally engineered by tailoring the shape, size, and
composition.
Scientists have approached the fabrication of quantum dots -- the ultimate in quantum confinement -- from
two very different points of view: (i) a top-down approach in which the extent and dimensionality of the
solid has gradually been reduced, and (ii) a bottom-up in which quantum dots are viewed as extremely large
molecules or colloids. The quantum dots grown by epitaxial and lithographic techniques are in the size
regime from 1 $\mu$m down to 10 nm, whereas the colloidal samples vary in diameter from the truly
molecular regime of 1 nm to about 20 nm. The latter systems are also known in the literature by the names
of colloidal quantum dots or nanocrystals: The spherical quantum dots (SQDs) are the precise examples of
this family playing better responsive role as lasers than their epitaxial-cum-lithographic counterparts [1].

The research interest burgeoned in SQDs has focused largely on the exciton dynamics of the bulk part with
the surface states, generally, eliminated by enclosure in a material of larger band-gap [2-7]. As such,
the scrutiny of the existing literature on the SQDs reveals the lack of genuine efforts devoted to
theorizing the magneto-optical absorption that rigorously justifies the (localized) plasmon excitation
peaks observed in the optical experiments. The present letter is motivated to fill that gap.

\section{Theoretical framework}
We consider a quasi-zero dimensional electron system (Q0DES) three-dimensionally confined by a harmonic
potential $V({\bf r})=\frac{1}{2}\,m^*\,\omega_0^2\, r^2$ and subjected to an applied magnetic field in
the symmetric gauge [${\bf A}({\bf r})=\frac{1}{2}\,({\bf B}\times {\bf r})$] in the spherical geometry
[with ${\bf r}\equiv (r, \theta, \phi)$]. For such a typical Q0DES,
the single-particle [of charge $-e$, with $e>0$] Hamiltonian in the Coulomb gauge
[$\nabla\cdot{\bf A}=0\Rightarrow {\bf A} \cdot {\bf p}={\bf p}\cdot {\bf A}$] can be expressed as
\begin{equation}
H = - \frac{\hbar^2}{2\,m^*}\nabla^2 + \frac{1}{2}\frac{e\,B}{m^*\,c}\,\hat{L}_z +
     \frac{1}{8}\frac{e^2}{m^*c^2}\,({\bf B}\times {\bf r})^2 + \frac{1}{2}m^*\,\omega_o^2\, r^2\, ,
\end{equation}
where $c$, $m^*$, ${\bf p}$, ${\bf A}$, ${\bf B}$, and $\omega_o$ are, respectively, the speed of light in
the vacuum, electron effective mass, momentum operator, vector potential, magnetic field, and the
characteristic frequency of the harmonic oscillator. The operator
$\hat{L}_z=-i\hbar \frac{\partial}{\partial \phi}$ is the z-component of the angular momentum. The harmonic
potential confining the SQDs is the most reasonable choice justifiable for the situation with small number
of electrons ($N$) in the dots. This model potential validates the generalized Kohn theorem (GKT) [8],
which states that the FIR resonant spectrum of a correlated many-electron system is insensitive to the
interaction effects. Obviously, we consider a one-component plasma inside the Q0DES and neglect the
spin-orbit interactions and the Zeeman energy for the sake of simplicity. Making use of the
Laplacian operator $\nabla^2$ (in the spherical coordinates), substituting
$\Psi(r,\theta,\phi)=R(r)\Theta(\theta)\Phi(\phi)$, and transposing allows us to solve the Schrodinger
equation $H\Psi=${\Large $\epsilon$}$\Psi$; with {\Large $\epsilon$} being the eigenenergy. The result is
that the Q0DES can be characterized by the eigenfunction
$\Psi(r,\theta,\phi)=R_{nl}(r)\,Y^m_l(\theta,\, \phi)$, where the radial function
\begin{equation}
R_{nl}(r)=N_r \, e^{-\mbox{\scriptsize X}/2} \, \mbox{\scriptsize X}^{l/2} \,
             \Phi \big(-\alpha_{nl};\, 1+\mbox{\large s};\,\mbox{\scriptsize X}\big)\, ,
\end{equation}
where $\mbox{\large s}=\frac{1}{2}+l$, $\mbox{\scriptsize X}=r^2/l^2_H$, and $N_r$ is the normalization
coefficient.
Here $\overbar{\mbox{\scriptsize X}}=\mbox{\scriptsize X}|_{r=R}$, and
$\Phi\big(-\alpha_{nl};\, 1+\mbox{\large s};\, \mbox{\scriptsize X}\big)$, $l_H=\sqrt{\hbar/(m^*\,\Omega_H)}$,
$\Omega_H=\sqrt{\frac{\omega^2_c}{4}\,\sigma^2+\omega^2_o}$, $\omega_c=eB/(m^*c)$, and $R$
are, respectively, the confluent hypergeometric function (CHF), the hybrid magnetic length, the hybrid
characteristic frequency, the cyclotron frequency, and the dot radius; and the spherical harmonics are defined
as
\begin{equation}
Y^m_l(\theta, \phi)=N_y\, P^m_l (\cos \theta)\, e^{i\,m\,\phi}\, ,
\end{equation}
where $P^m_l (\cos \theta)$ is the associated Legendre function and $N_y$ is the normalization coefficient.
Here $n$, $l$, and $m$ are,
respectively, the principal, orbital, and magnetic quantum numbers. The spherical harmonics satisfy the
well-knwon condition of orthonormality.
The finiteness of the quantum dot requires that the eigenfunction $\Psi (...)$ satisfies the boundary
condition
$R_{nl} (r=R)=0\Rightarrow \Phi(-\alpha_{nl};\, 1+\mbox{\large s};\, \overbar{\mbox{\scriptsize X}})=0$.
This determines the eigenenergy of the system defined by
\begin{equation}
\mbox {\Large $\epsilon$}_{nlm}=2\,\hbar\Omega_H\Big[\alpha_{nl}+
                 \frac{1}{2}\,\big(1+ \mbox{\large s}\big)\Big]+
                     \frac{1}{2}\,m\,\hbar\omega_c
\end{equation}
In this form, the eigenenergy for SQDs is {\em formally} identical to that for the quantum dots laterally
confined in the Q2DES [1]. It is important to notice the appearance of the symbol $\sigma$ defined by
$\sigma^2= \big<l'm'\big|\sin^2\theta \big|lm \big>$ in the definition of $\Omega_H$. In order to
investigate the magneto-optical absorption in the semiconducting SQDs, what we actually need to compute
sagaciously is only the energy dependence of the {\em imaginary} part of the interacting (or total)
density-density correlation function (DDCF) $\chi (...)$, where
\begin{align}
\chi(\omega)
=&\int d{\bf r}\int d{\bf r}'\,\chi({\bf r},{\bf r}';\omega)\nonumber\\
=& \sum_{ijkl}\,\Big[\Pi_{ij}\,\delta_{ik}\,\delta_{jl}+\Pi_{ij}\,\sum_{mn}\,
                      \Lambda_{klmn}\,\Pi_{mn}\,F_{mnij}\Big]\,\nonumber\\
 & \hspace {0.63cm} \times \int d{\bf r}\int d{\bf r}'\,
                      \Psi^*_i({\bf r})\Psi_j({\bf r})\Psi^*_l({\bf r}')\Psi_k({\bf r}')\, ,
\end{align}
where $\Pi_{ij}=[f(\epsilon_i)-f(\epsilon_j)]/[\epsilon_i-\epsilon_j+\hbar\,\omega^+]$, $\Lambda_{\mu\nu}$
is the inverse of $[\delta_{\mu\nu}-\Pi_{\mu}\beta_{\mu\nu}]$ such that $\sum_{\mu}\,\Lambda_{\gamma\mu}\,[\delta_{\mu\nu}-\Pi_{\mu}\beta_{\mu\nu}]=\delta_{\gamma\nu}$, and
$F_{mnij}=\int d{\bf r}''\int d{\bf r}'''\,
                  \Psi^*_m({\bf r}'')\Psi_n({\bf r}'')\,V_{ee}({\bf r}'',{\bf r}''')\,
                  \Psi^*_j({\bf r}''')\Psi_i({\bf r}''')$ stands for the matrix elements of the Coulombic
interaction potential $V_{ee}({\bf r}, {\bf r}')$. Here $f(\epsilon_j)$ is the Fermi-Dirac distribution
function, and $\beta_{\mu\nu}=\int d{\bf r}\, L^*_{\mu}({\bf r})\,S_{\nu}({\bf r})$; with the symbols
$L_{\mu}({\bf r})$ and $S_{\mu}({\bf r})$ representing, respectively, the long-range and the short-range
parts of the Coulombic interactions discussed in the strategy leading to the determination of the inverse
dielectric function $\epsilon^{-1}({\bf r},{\bf r}';\omega)$ [1] and the subscript $\mu,\,\nu\equiv i,\,\,j$
is a composite index introduced just for the sake of the mathematical convenience. Equation (5) is a variant
of the Dyson equation derived within the framework of the Bohm-Pines' full RPA [1].
\begin{figure}[htbp]
\includegraphics*[width=7.5cm,height=6.5cm]{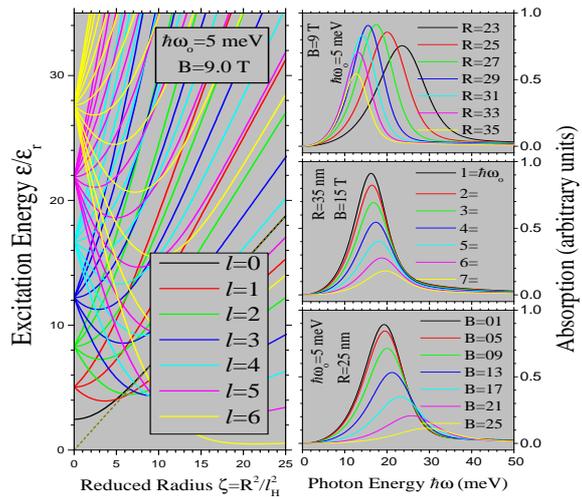}
\caption{(Color online) Left panel: The Fock-Darwin excitation spectrum for the GaAs SQDs in the presence of
an applied magnetic field ($B$), for several values of  $l\le 6$. The dashed line
represents the lowest branch of the magnetic (or Landau) fan -- when $\alpha_{nl}\rightarrow 0$ for very
large $R$ or $B$. We chose not to plot the whole magnetic fan in order to avoid the mess.}
\caption{(Color online) Right panel: The magneto-optical absorption vs. the excitation energy for the GaAs
SQDs for several values of the dot radius $R$ [top panel], confinement potential $\hbar\omega_o$ [middle panel],
and magnetic field $B$ [bottom panel]. Other parameters are listed inside.}
\label{fig1}
\end{figure}

\section{Results and Discussion}
We focus on colloidally prepared GaAs SQDs, which implies that the background dielectric constant
$\epsilon_o=12.8$ and $m^*=0.067\,m_o$; with $m_o$ as the bare electron mass. The other parameters involved
in the computation are: $R$, $\hbar\omega_o$, and $B$. Until and unless stated otherwise, we specify the
quantum numbers such that the dot contains (in total) 24 electrons (including spin). Also, we assume
compliance of the lowest subband approximation due to harmonic confinement, which makes sense for the dots
with small charge densities at low temperatures where most of the experiments are performed on the
low-dimensional systems.

Figure 1 represents the Fock-Darwin excitation spectrum for the spherical quantum dot in the presence of a
confinement potential ($\hbar\omega_o=5$ meV) and an applied magnetic field ($B=9$ T) for the orbital quantum
number $l\le 6$. The plots are rendered in terms of the dimensionless energy $\epsilon_{nlm}/\epsilon_r$
and the dot radius $\zeta=R^2/l^2_H$; with $\epsilon_r=2\,\hbar^2/(m^*\,R^2)$. First and foremost, we observe
that, unlike the laterally confined 2D quantum dots [1], the electronic levels at the origin are not equispaced.
The ($2\,l+1$)-degeneracy at the origin justifies the intuition. All the modes with $m\ge 0$ are seen to start
and propagate throughout with a positive group velocity, whereas those with $m<0$ start with a negative group
velocity, attain a minimum as $\zeta$ increases, and finally propagate with the positive group velocity. The
larger the $|m|$, the greater the value of $\zeta$ where the latter type of modes observe a minimum. The dashed
line starting from zero and finally merging with ($l=0=m$) mode is the lowest branch of the magnetic fan. At
large values of $B$, $\zeta$ becomes large, the role of $\alpha_{nl}$ diminishes, and the magnetic fan is born.

Figure 2 illustrates the magneto-optical absorption as a function of the excitation energy for several values
of the dot radius [top panel], confinement potential $\hbar\omega_o$ [middle panel], and magnetic field $B$
[bottom panel]. It is observed that the absorption peak experiences a blue shift as the confinement potential
increases, the dot radius decreases, and the magnetic field increases. For example, by decreasing the radius
from $R=35$ nm to $R=23$ nm, the peak position has blue-shifted by $31\%$. Similarly, increasing the confinement
from $\hbar\omega_o=1$ meV to $\hbar\omega_o=7$ meV results in the peak position blue-shifted by $21\%$. Again,
increasing the magnetic field from $B=1$ T to $B=25$ T yields a peak position blue-shifted by $52\%$. To sum
up, reducing the dot-size and intensifying the confinement or the magnetic field enhances the capacity of the
quantum dots to absorb the photons of higher energy. Also, the FWHM of the absorption peak increases with
decreasing dot size and with increasing confinement or magnetic field. This reflects the tendency of
the oscillator strength focused into a fewer transitions for the stronger confinement. By and large, the size
effects predominate over those due to confinement or magnetic field.

Figure 3 shows the radial probability distribution (RPD) in the ground state against the (reduced) radial
coordinate $r/R$. The RPD $Q_{nl}$($r$) defined as
$Q_{nl}(r)=\int^{\pi}_{0} d\theta\,\sin\theta\,\int^{2\pi}_0 d\phi\,r^2\,|\Psi(r, \theta, \phi)|^2$ expresses
the probability of finding an electron as a function of $r$ at a given instant. The upper panel plots the
ground states ($l=0=m$) along with $n=1$, 2, 3, 4, and 5 for the magnetic field $B=15$ T. In the language
of the {\em spectroscopy}, the principal quantum number $n$ distinguishes the $s$ orbitals -- characterized by
$l=0$. 
The lower panel of Fig. 3 exhibits the RPD in the ground state
for several values of $B$ in the range $0\le B$\,(T)\,$\le15$. Note that $Q_{nl}$($r$) is zero at $r=0$ because
the volume of space available $r^2\,dr=0$. As $r$ increases, the dot size increases and so does $Q_{nl}$($r$).
However, the probability density $|\Psi(r, \theta, \phi)|^2$ decreases with increasing $r$, which implies that
the $Q_{nl}$($r$)-path must have observed a maximum, where the probability of locating the electron is eminent.
The influence of $B$ on the variation of $Q_{nl}$($r$) is remarkable: while the magnetic field tends to further
add to the confinement and hence maximize the probability distribution with increasing $B$, it
does not shift the peak position of the radial distribution, which lies at $r/R\simeq 0.495$ for the whole range
of $B$, including $B=0$.

Figure 4 describes the magneto-optical transitions in the spherical quantum dots as a function of magnetic
field for a given value of the dot size ($R=35$ nm), for several values of the confinement potential
($05\le \hbar\omega_o$ (meV) $\le 15$). Figure 4 is based on the computation of the exact analytical
expression for the transition energy:
$\Delta{\mbox{\Large $\epsilon$}}=2\,\big[\Delta\alpha\,\pm\,\frac{1}{2}\big]\,\hbar\Omega_H\,
                                      \pm \,\frac{1}{2}\,\hbar\omega_c$;
where $\Delta\alpha\equiv \Delta\alpha_{nl}=\big[\alpha_{1,1}-\alpha_{1,0}\big]$.
Some simple mathematical
manipulations of this expression reveal that, unlike the laterally confined quantum dots [1], both
-- upper [$\Delta\epsilon_+$] and lower [$\Delta\epsilon_-$] -- transitions survive whether or not
$\hbar\omega_o=0$. Similarly, at $B=0$, we obtain two, albeit relatively weaker, transitions
given by $\Delta\epsilon_{\pm}=2 [\Delta\alpha \pm \frac{1}{2}]\,\hbar\omega_o$.
Even when $B\rightarrow\infty$ [i.e., $\omega_c\gg\omega_o$], we cannot avoid either of the two transitions.
In other words, the SQDs do not allow the intra-Landau level transitions. Moreover, the non-zero
parameters $\sigma$ and $\alpha_{nl}$ prevent the edges of the wedges to be exactly characterized by the
confinement potential ($\hbar\omega_o$) at $B=0$. These findings should prompt interesting magneto-optical
experiments on the SQDs aimed at verifying such predictions.
\begin{figure}[htbp]
\includegraphics*[width=7.5cm,height=6.5cm]{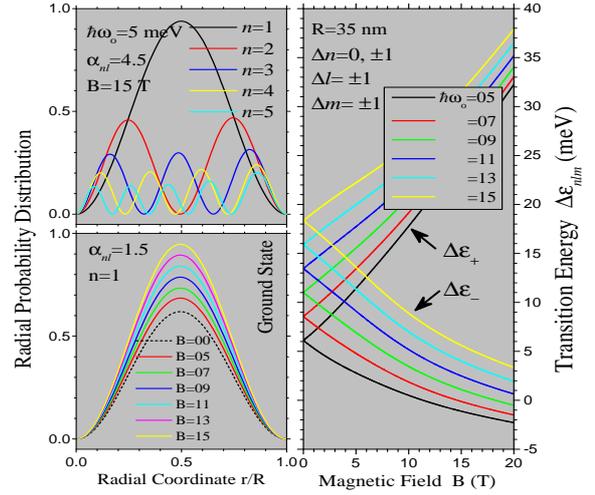}
\caption{(Color online) Left panel: The radial probability distribution $Q_{nl}$($r$) in the ground state for
the GaAs SQDs against the reduced (radial) coordinate $r/R$, for several values of $n$, which distinguishes
the $s$ orbitals characterized by $l=0$. The number of nodes equals ($n-1$). The lower panel
shows $Q_{nl}$($r$) in the ground state ($n=1,\, l=0,\, m=0$) for several values of $B$.}
\caption{(Color online) Right panel: Allowed magneto-optical transitions vs. the magnetic field
($B$), for several values of the confinement potential $\hbar\omega_o$. The dot radius is defined as $R=35$ nm.
The selection rules are listed in the picture. The upper (lower) transition is designated as $\Delta\epsilon_+$
($\Delta\epsilon_-$).}
\label{fig3}
\end{figure}
\section{Concluding remarks}
The magneto-optical absorption peak is blue-shifted with reducing dot-size and intensifying confinement or
magnetic field. However, the size effects are observed to be predominant. While the magnetic field tends
to maximize the probability of finding the electron, it does not shift the peak position of the RPD.
Both -- upper and lower -- magneto-optical transitions survive even in the extreme cases of vanishing
confinement or magnetic field. A deeper grasp of the importance of the spherical geometry [9] and of related
aspects such as the quantum coherence, Coulomb blockade, and entanglement may lead to a better insight into
the promising applications involving lasers, detectors, storage devices, and quantum computing, to name a few.
\vskip3pt
\ack{Sincere thanks to Loren Pfeiffer, Daniel Gammon, and Aron Pinczuk for the useful communication.}

\vskip5pt

\noindent Manvir S. Kushwaha
(\textit{Rice University, P.O. Box 1892, Houston, TX 77251})
\vskip3pt

\noindent E-mail: manvir@rice.edu


\begin{thebibliography}{}


\bibitem{1} For an extensive review of electronic, optical, and transport phenomena in the systems
            of reduced dimensions such as quantum wells, quantum wires, and quantum dots,
            see M.S. Kushwaha, 'Plasmons and magnetoplasmons in semiconductor heterostructures',
            Surf. Sci. Rep., 2001, {\bf 41}, pp. 1-416
\bibitem{2} A.L. Efros and A.L. Efros, 'Interband absorption of light in a semiconductor sphere', Sov.
            Phys.: Semiconductors, 1982, {\bf 16}, pp. 772-775
\bibitem{3} L.E. Brus, 'Electron-electron and electron-hole interactions in small semiconductor
            crystallites: The size dependence of the lowest excited electronic state', J. Chem. Phys.,
            1984, {\bf 80}, pp. 4403-4409
\bibitem{4} L. Banyai and S.W. Koch, 'Absorption blue shift in laser-excited semiconductor microspheres',
            Phys. Rev. Lett., 1986, {\bf 57}, pp. 2722-2724
\bibitem{5} M.L. Steigerwald and L.E. Brus, 'Synthesis, stabilization, and electronic structure of
            quantum semiconductor nanocrystals', Annu. Rev. Mater. Sci., 1989, {\bf 19}, pp. 471-495
\bibitem{6} Y. Wang and N. Herron, 'Nanometer-sized semiconductor clusters: Materials synthesis,
            quantum size Effects, and photophysical properties', J. Phys. Chem., 1991, {\bf 95},
            pp. 525-532
\bibitem{7} A.P. Alivisatos, 'Semiconductor clusters, nanocrystals, and quantum dots', Science, 1996,
            {\bf 271}, pp. 933-937
\bibitem{8} F. M. Peeters, 'Magneto-optics in parabolic quantum dots', Phys. Rev. B, 1990, {\bf 42},
            pp. 1486-1487
\bibitem{9} J.K. Jain, {\it Composite Fermions} (Cambridge, New York, 2007).
\end{thebibliography}
\end{document}